\documentclass[pre,onecolumn]{revtex4}
\usepackage {graphicx}

\begin{document}
\title{Autocatalytic reaction-diffusion processes in restricted geometries}

\author{Elena Agliari}
\email{elena.agliari@fis.unipr.it}
\author{Raffaella Burioni, Davide Cassi, Franco M. Neri}

\affiliation{
Dipartimento di Fisica, Universit\`{a} degli Studi di Parma, Parco Area
delle Scienze 7/A, 43100 Parma, Italy}

\begin{abstract}
We study the dynamics of a system made up of particles of two different species undergoing irreversible quadratic autocatalytic reactions:
$A + B \rightarrow 2A$. We especially focus on the reaction velocity and on the average time at which the system achieves its inert state. By means of both analytical and numerical methods, we are also able to highlight the role of topology in the temporal evolution
of the system.
\end{abstract}

\maketitle

\section{Introduction}

The interest in systems undergoing reaction-diffusion processes is
experiencing a rapid growth, due to their intrinsic relevance in an
extraordinary broad range of fields [1].

In particular, a great deal of experimental and theoretical work has been
devoted to the study of reaction-diffusion processes embedded in \textit{restricted geometries}. This
expression refers to two, possibly concurrent, situations: $i.$ low
dimensionality and \textit{ii.} small spatial extent.

In the first case, the spectral dimension $\tilde{d}$ characterizing the
diffusive behaviour of the reactants on the substrate is low $(1 < \tilde{d} <
2)$, and the substrate underlying the diffusion-reaction lacks spatial
homogeneity. This situation is able to model media whose properties are not
translationally invariant and where the reactants perform a ``compact
exploration'' [2]. These kinds of structures can lead to a chemical
behaviour significantly different from those occurring on substrates
displaying a homogeneous spatial arrangement. Indeed, while in high
dimensions a mean-field approach (based on classical rate equations)
provides a good description, in low dimension local fluctuations are
responsible for significant deviation from mean-field predictions [3].

There also exists a variety of experimental situations in which
reaction-diffusion processes occur on spatial scales too small to allow an
infinite volume treatment: in this case finite-size corrections to the
asymptotic (infinite-volume) behaviour become predominant.

Here, differently from previous works, we explicitly examine finite size
systems, i.e. no thermodynamic limit is taken [4-6]. All the quantities we
calculate are hence finite, and we seek their dependence on the finite
parameters of the system (volume of the substrate and concentration of the
reactants). In particular, we study the dynamics of a system made up of two
species particles undergoing irreversible quadratic autocatalytic reactions
$A + B \rightarrow 2A$. All particles move randomly and react upon encounter with
probability $1$, i.e. the reaction is strictly local and
deterministic. Notice that, allowing all the particles to diffuse makes the
problem under study a genuine multiparticle-diffusion problem. The latter is
generally quite difficult to manage due to the fact that the effects of each
single particle do not combine linearly, even in the non-interacting case.
For this reason the analytic treatment often relies on simplifying
assumptions which, nevertheless, preserve the main generic features of the
problem. In the past, autocatalytic reactions have been extensively analyzed
on Euclidean structures [7], within a continuous picture attained by the
Fisher equation [8,9] which describes the system in terms of front
propagation. Evidently, this picture is not suitable for low-density
systems, where front propagation cannot be defined. In order to describe
also the high-dilution regime, here a different approach is introduced
which, as we will see, works as well for inhomogeneous structures. This way,
we are also able to highlight the role of topology in the temporal evolution
of the system.

In the following, we shall examine the concentration $\rho _{A}(t)$ of $A$
particles present in the system at time $t$ and its fluctuations; from $\rho
_A(t)$ it is then possible to derive an estimate for the reaction
velocity. Furthermore, we consider the average time $\tau $ (also called
``Final Time'') at which the system achieves its inert state, i.e.
$N_{A}=N$. As we will show, $\tau$ depends on the number of particles $N$ and
on the volume $V$ of the underlying structure. More precisely, for small
concentrations of the reactants, we find, both numerically and analytically,
that the $\tau$ factorizes into two terms depending on $N$ and $V$,
respectively.

One of the most interesting applications of the Final Time is analytic
[10,11]: as we show, $\tau $ sensitively depends on the initial amount of
reactant $N$ and, on low dimensional substrates $(d < 2)$, by reducing the
dimension $d$, the sensitivity can be further improved.

\section{The model}
We consider a system made up of $N$ particles of two different chemical
species $A$ and $B$, diffusing and reacting on a discrete substrate with no
excluded volume effects. At time $t$, $N_{A}(t)$ and $N_{B}(t)$ represent the
number of $A$ and $B$ particles, respectively, with $N=N_{A}+N_{B}$. Being $V$
the substrate volume, we define $\rho _{A}(t)=N_{A}(t)/V$ and $\rho_{B}(t)=N_{B}(t)/V$ as the concentrations of the two species at time $t$.

Different species particles residing at time step $t$, on the same node or on
nearest-neighbour nodes react according to the mechanism $A + B \rightarrow 2A$
with reaction probability set equal to one. Notice that the previous scheme
also includes possible additional products (other than $2A$) made up of some
inert species of no consequences to the overall kinetics. The initial
condition at time $t=0$ is $N_{A}(0)=1$ (the Source), $N_{B}(0)=N-1$, with all
particles distributed randomly throughout the substrate. As a consequence of
the chemical reaction defined above, $N_{A}(t)$ is a monotonic function of $t$
and, due to the finiteness of the system, it finally reaches value $N$; at
that stage the system is chemically inert. The average time at which
$N_{A}(t)=N$ is called ``Final Time'' and denoted by $\tau$.

The Final Time $\tau$ is of great experimental importance since it
represents the average time when the system is inert and therefore it
provides an estimate of the time when reaction-induced effects (such as
side-reactions or photoemission) vanish [12]. In this perspective,
deviations from the theoretical prediction of $\tau$ are, as well,
noteworthy: they could reveal the existence of competitive reactions or
explain how the process is affected by external radiation.

Finally, notice that the autocatalytic reaction can also be used as a model
for spreading phenomena: $A (B)$ particles may stand for (irreversibly) sick
(healthy) or informed (unaware) agents, respectively. For these systems a
knowledge of the infection rate or information diffusion is of great
importance [4,5].

\section{Average Final Time}
As previously said, $\tau$ generally depends on the total number of agents
$N$ and on the size of the lattice $V$, while its functional form is affected by
the topology of the lattice itself. The analytical treatment is carried out
in the two limit regimes of high and low density.

\subsection{High-density regime}
When $\rho =N/V \gg 1$, the substrate topology does not qualitatively affects
results. We can assume that the set of $A$ particles covers a connected region
of the substrate whose volume expands with a constant velocity (depending on
the density $\rho$ and dimension $d$). In this case (and exactly in the limit
$\rho  \to \infty )$ the process can be described as the deterministic
propagation of a wave front decoupled from the random motion of the agents.
If we suppose the Source to be at the center of the lattice at time $t=0$, at
each instant the wave front is the locus of points whose chemical distance
from the center is $2t+1$. The connected region spanned by the wave front is
entirely occupied by $A$ particles, while $B$ particles fill the remaining of
the lattice. In particular, for a $d$-dimensional regular substrate, the
region where $A$ particles concentrate is a $d$-dimensional polyhedron [4,5].

In general, for a finite system, the average Final Time is $\tau
=l_\mathrm{{max}}/2$, where $l_{\mathrm{max}}$ is the chemical distance of the most distant
point on the lattice, starting from the Source. On Euclidean geometries this
yields $\tau=L/4$ for d=1 and $\tau=L/2$ for $d \ge 2$. On the other hand,
on inhomogeneous structures, the dependence on $L$ is not so simple, since it
involves taking the average with respect to all possible starting points for
the Source.

\subsection{Low-density regime}
In the case of low density $(\rho \ll 1)$ the time an $A$ particle walks before
meeting a $B$ particle becomes very large, so that the process is
diffusion-limited. We adopt a mean-field-like approximation by assuming that
the time elapsing between a reaction and the successive one is long enough
that the spatial distribution of reactants can be considered random. In
other words, the particles between each event have the time to redistribute
randomly on the lattice and we neglect correlations between their spatial
positions. Another consequence of the low concentration of reactants, is
that we can just focus on two-body interactions since the event of three or
more particles interacting together is unlikely. Notice that the
high-dilution assumption, by itself, generally does not allow to apply the
classical rate equations: when diffusion is involved also the substrate
topology has to be taken into account. For this reason, in the following we
will treat high and low dimensional structures separately.

\bigskip

\textit{High-dimensional structures} $(\tilde{d}>2)$ Let us consider a given configuration of the system where
$N_{A}$ and $N_{B}$ particles are present. The probability for a given B
particle to encounter and react with any A particle is just the trapping
probability $P_{trap}(\rho _{A},t)$ for a particle, out of $N_{B}$, in
the presence of $N_{A}$ traps, both species diffusing. Under the assumptions
specified above, for high-dimensional substrates [1]:
\begin{equation}
P_\mathrm{{trap}}(\rho _{A},t) = \lambda _{d} \; \rho_{A} e^{-\lambda_d \rho_A t},
\end{equation}
where $\lambda_{d}$ is a constant depending on the given substrate. Form
the previous equation we can calculate the average trapping time for a B
particle as $\tau_\mathrm{{trap}}(\rho_{A}) \; \sim  \; \rho_{A}^{-1}$.

Let us now introduce an early-time $(t < \tau_\mathrm{{trap}}(\rho_{A}))$
approximation for the trapping probability: $P_\mathrm{{trap}}(\rho_{A},t) \sim
p N_{A}$, where $p \sim V^{ - 1}$ is the probability that, after each
reaction, two given particles first encounter at a given time (in general,
this probability depends not only on the volume of the underlying structure,
but also on the history of the system). This simple form for
$P_\mathrm{{trap}}(\rho_{A},t)$ allows us to go on straightforwardly. In fact,
the process can be meant as an absorbing Markov chain, with $N$ states
(labeled with the total number of $A$ particles: $1,2,...,N$), and one
absorbing state $(N_{A}=N)$; the chain starts from state 1. The transition
matrix $\mathbf{P}$ can be written: the transition probability from a state
$k$ to a state $m$ as a function of N and p is:
\begin{equation}
P_{k,m} = {N - k \choose m - k}\left[ 1 - (1 - p)^k \right]^{m - k}\left[ (1 -
p)^k \right]^{N - m}
\end{equation}
for any $N$ and $p$. From $\mathbf{P}$ we can take the submatrix $\mathbf{Q}$,
obtained subtracting the last row and column (those pertaining to the
absorbing state), and compute the fundamental matrix
$\mathbf{F}=(1-\mathbf{Q})^{-1}$. Now, by expanding to first order in $p$, a
direct calculation shows that $\mathbf{F}$ is an upper triangular matrix given
by
\begin{equation}
F_{k,m}=
\left\{
\begin{array}{cl}
\frac{1}{m\left( {N - m} \right)p} &  k \ge m  \\
0 & k < m.
\end{array}
\right.
\end{equation}
The mean time $\tau$ required to reach the absorbing state N, starting from
state 1 is given by the sum of the first row of $\mathbf{F}$:
\begin{equation}
\label{eq1}
\tau \left( {N,V} \right) = \frac{1}{p}\sum\limits_{m = 1}^{N - 1}
{\frac{1}{m(N - m)}\mathrel{\mathop{\kern0pt\longrightarrow}\limits_{N \to
\infty }} \frac{V\left( {\gamma + \log N} \right)}{N}}
\end{equation}
where $\gamma =0.577...$ is the Euler-Mascheroni constant. The last result
is in perfect agreement with numerical simulations and also emphasizes $\tau$ factorization.

\bigskip

\textit{Low-dimensional structures} $(\tilde{d} \le 2)$. For low dimensional structures the dependence on $N$
found above is not correct. The reason is that a non-linear cooperative
behaviour among particles emerges.

Let us define $\langle t_{n} \rangle$ as the average time elapsing between the $(n-1)$-th
first encounter among different particles and the $n$-th one. This time just
corresponds to the average time during which there are just $N_{A}(t)=n$
particles in the system. In our approximation $\langle t_{n} \rangle$ is proportional to
the trapping time $\tau_\mathrm{{trap}}(n/V)$ in the presence of n mobile traps
diffusing throughout a volume $V$ [6]. For compact exploration of the space
$(\tilde{d}<2)$, $\tau_\mathrm{{trap}}(n/V)\sim (V/n)^{2 / \tilde{d}}$. This result was
derived for infinite lattices, nonetheless, it provides a good approximation
also for finite lattices, provided that the time to encounter is not too
large. From $\tau_\mathrm{{trap}}$ we obtain $\langle t_{n} \rangle$ as the average trapping
time of the first out of $N-n \equiv N_{B}$ particles, that, for rare
events, is just
\begin{equation}
\label{eq2}
\langle {t_n} \rangle = V^{2 / \tilde{d} }\frac{n^{ - 2 / \tilde{d} }}{N -
n},
\end{equation}
with logarithmic corrections in the case $\tilde{d}=2$. The time $\tau$ can
therefore be written as a sum over $n$ $(n=1, 2, \ldots , N-1)$ of $\langle t_{n} \rangle$.
Now, by adopting a continuous approximation, we obtain for $\tilde{d}<2$ [6]:
\begin{equation}
\label{eq3}
\tau (N,V) \approx V^{2 / \tilde{d} }\left[ {\frac{\tilde{d} }{(2 - \tilde{d} )N} + N^{ - 2 /
\tilde{d} }\left( {\log N + H_{2 / \tilde{d} } } \right) + O(N^{ - 1})} \right],
\end{equation}
where $H_{m}$ is the harmonic number. In particular, the leading-order
contribution for a one-dimensional system $(\tilde{d}=d=1)$ is
\begin{equation}
\tau (N,V) \approx \frac{V^2}{N}.
\end{equation}

For a two-dimensional lattice $(\tilde{d}=d=2)$
\begin{equation}
\label{eq4}
\tau (N,V) \approx V\log V\frac{\log N + \gamma }{N}.
\end{equation}
Notice that the factorization in Eq.(\ref{eq3}) is consistent with Eq.(\ref{eq1}): in both
cases, the factor containing the dependence on $V$ represents the average time
for two particles to meet.
\begin{figure}[htbp]\label{fig1}
\centerline{\includegraphics[width=3.0in,height=2.8in]{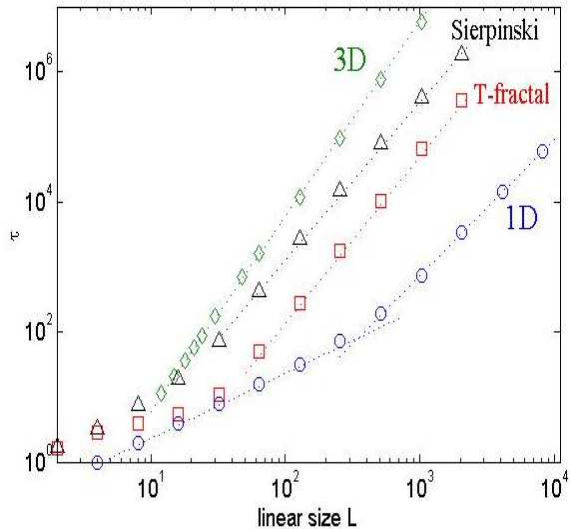}}
\caption{Scaling of $\tau$ with the linear size of the system for a
one-dimensional chain (blue circles), a Sierpinski gasket (black triangles),
a T-fractal (red squares), and a three-dimensional cubic lattice (green
diamonds) on a double-logarithmic scale. The number of reactants is fixed at
$N=1024$ for all systems. The spectral dimension for the Sierpinski and the
T-graph is $\tilde{d} \approx $1.365 and $\tilde{d} \approx 1.226$,
respectively. Dotted lines highlight the low-concentration regime $(L \gg 1)$,
corresponding to a power law for all systems. For the one-dimensional chain,
the linear high-concentration regime is also pointed up.}
\end{figure}

As can be evinced from Fig. 2, for small densities all the data collapse;
moreover, in that region, the fit coefficients introduced are in good
agreement with theoretical predictions.

\begin{figure}[htbp]\label{fig2}
\includegraphics[width=2.4in,height=2.00in]{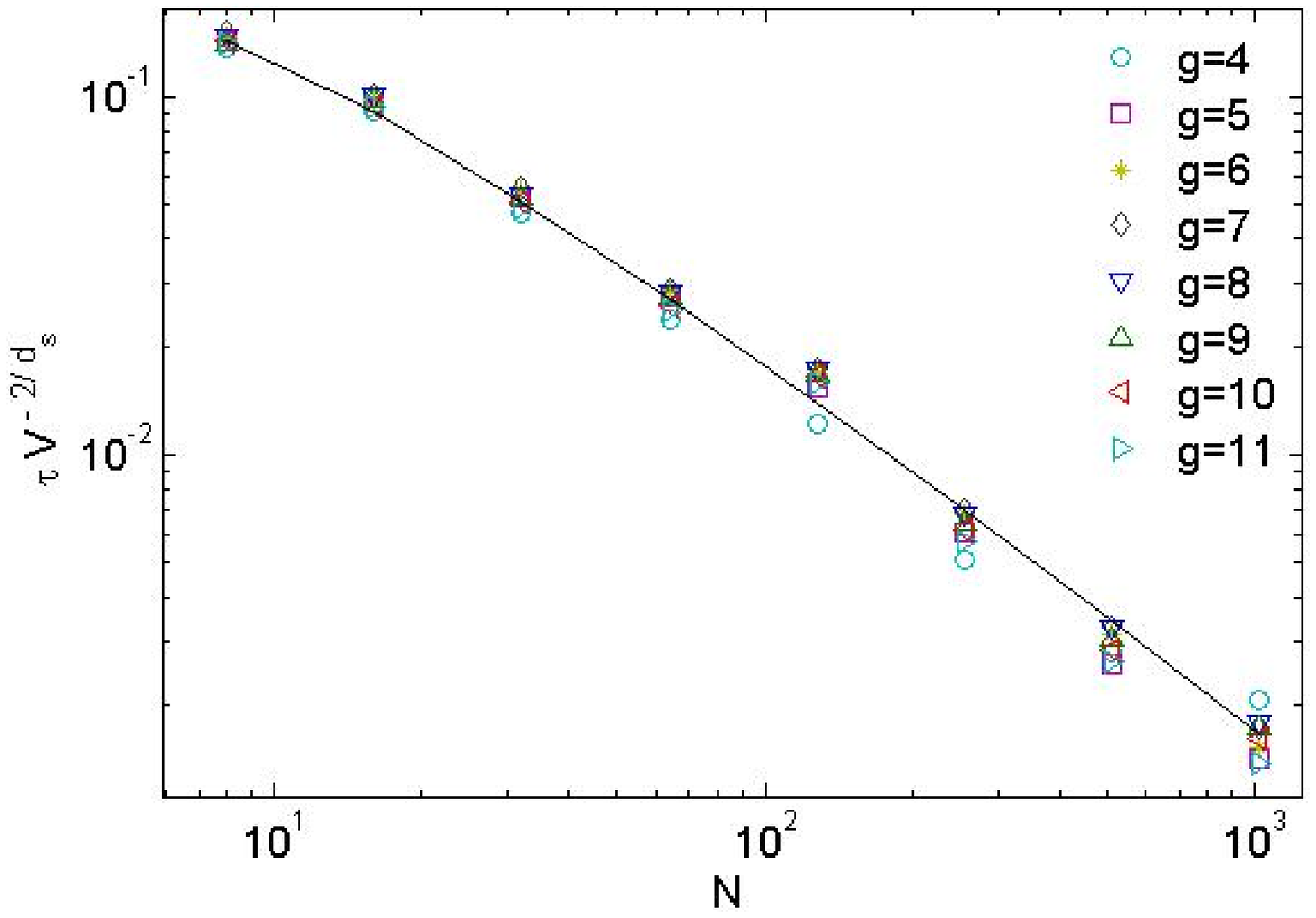}
\includegraphics[width=2.4in,height=2.05in]{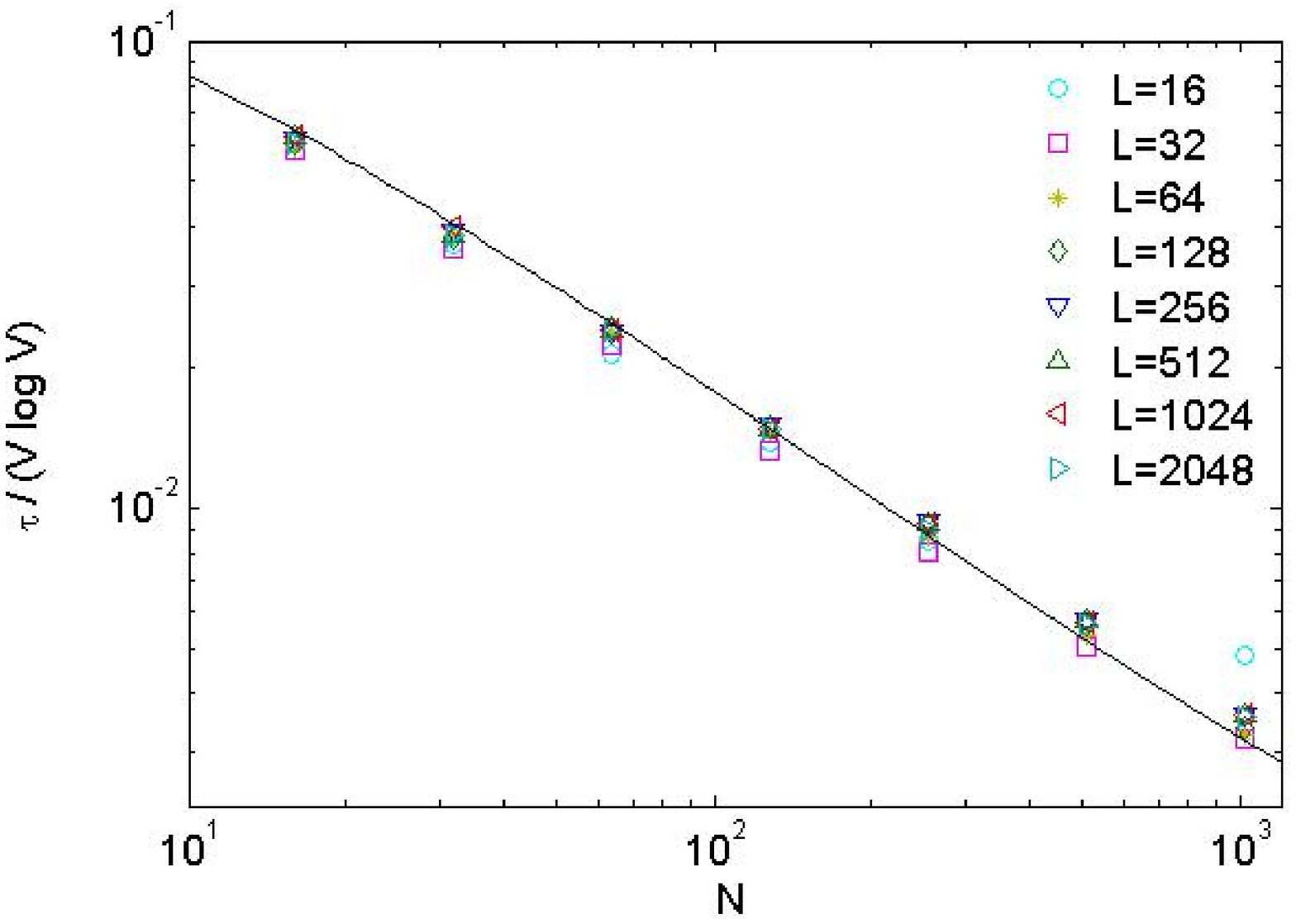}
\caption{Rescaled Final time versus number of particles $N$ for the Sierpinski
gasket (left) and the two-dimensional lattice (right). Different symbols and
colours distinguish different sizes, as explained by the legend. The line
provides the best fit in very good agreement with Eqs. (\ref{eq3}) and (\ref{eq4}), apart
from sub-leading corrections in the (marginal) case $d=2$.}
\end{figure}

For low densities, the standard deviation $\sigma _{\tau}$ displays a
dependence on $N$ and $L$ analogous to $\tau$; for high densities, $\sigma_{\tau}$ becomes vanishingly small, in fact the process becomes
deterministic.

As anticipated in Section 1, experimental measures of $\tau $ are useful in
monitoring trace reactants [6]. In the high-dilution regime, our results
show that $\tau =f_{\tilde{d}}(N)g_{\tilde{d}}(V)$ and therefore, once the substrate
size is fixed, the initial amount of reactant can be expressed as $N =
f_{\tilde{d}}^{-1}(\tau / g_{\tilde{d}}(V))$.

A proper estimate of the sensitivity of this method is provided by the
derivative $\partial N/ \partial \tau $: the smaller the derivative and
the larger the sensitivity. As can be evinced from Fig.3, which displays
numerical results for $N$ and $\partial N/ \partial \tau$, the smaller the
concentration and the better the sensitivity of this technique. This makes
such technique very suitable for the determination of ultratrace amounts of
reactants, which is of great experimental importance [13]. Interestingly,
$\partial N/ \partial \tau$ also depends on the substrate topology: when
$\tilde{d} \le $ 2 and at fixed $V$, the sensitivity can be further improved by
lowering the substrates dimension. Conversely, when $\tilde{d} > 2, \partial
N/\partial \tau$ ceases to depend on $\tilde{d}$.

\begin{figure}[htbp]\label{fig3}
\includegraphics[width=2.40in,height=2.00in]{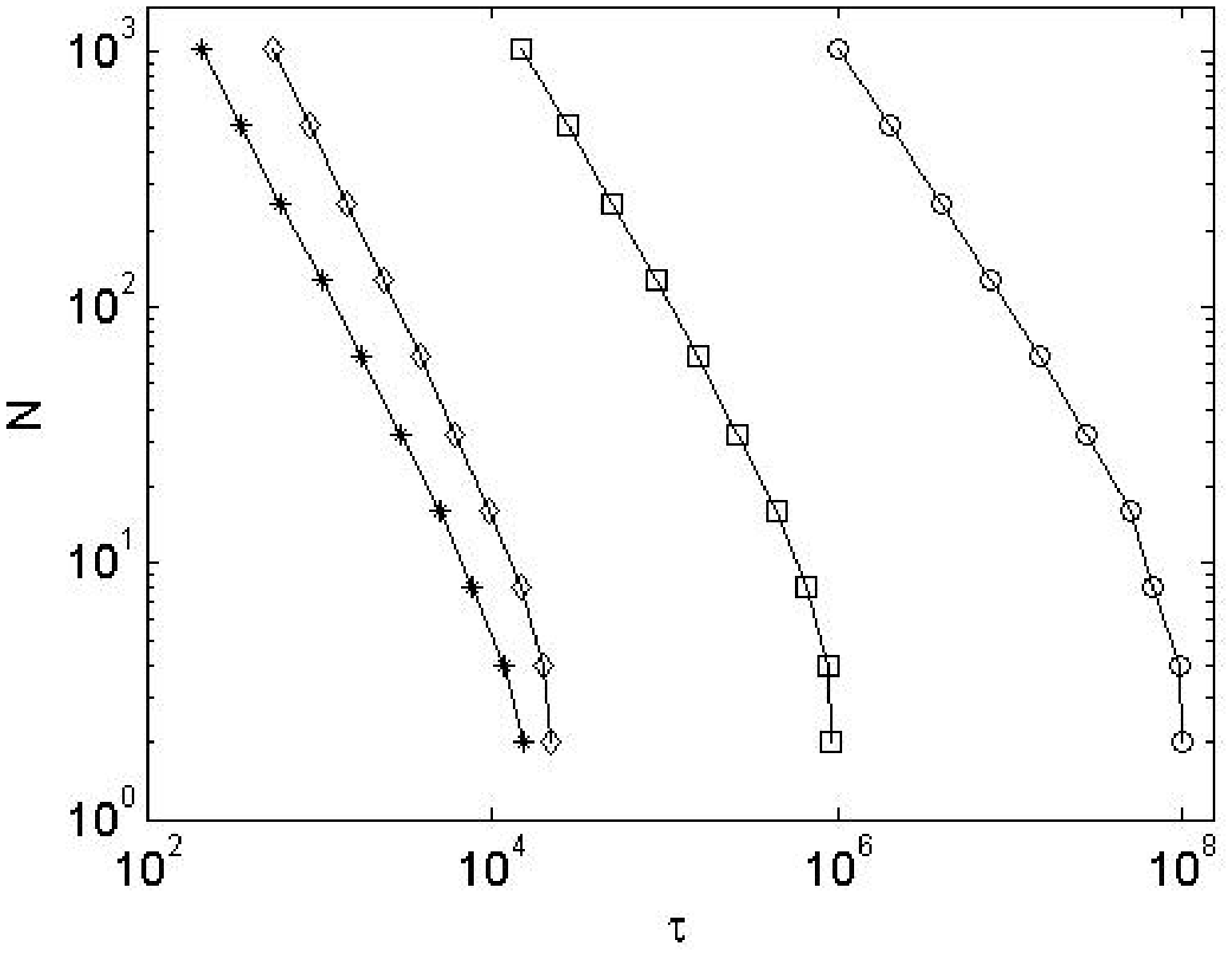}
\includegraphics[width=2.40in,height=2.00in]{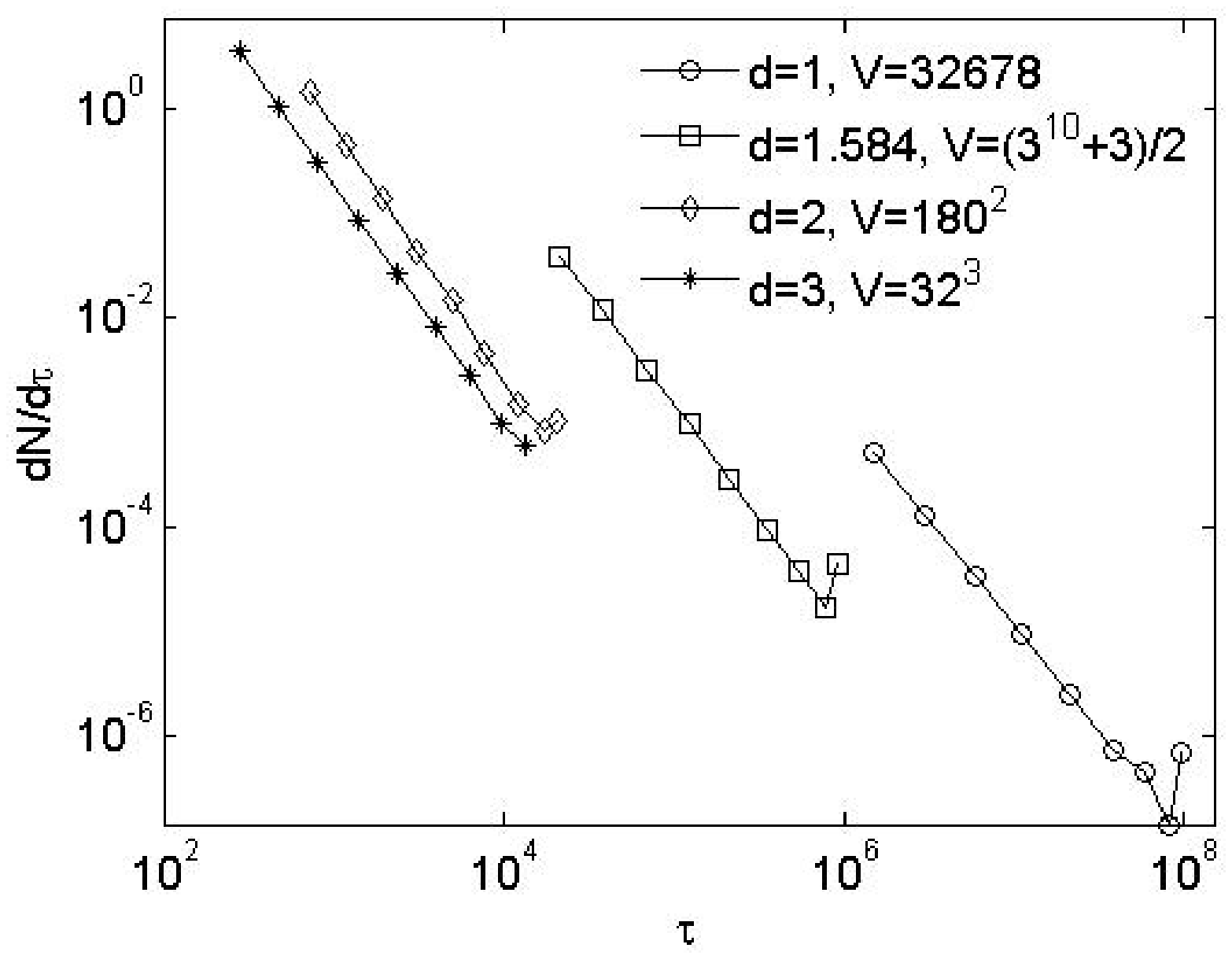}
\caption{Log-log scale plot of the reactant amount $N$ (left panel) and its
derivative $\partial N/ \partial \tau$ (right panel) vs Final Time $\tau$. As shown in the
legend, different substrate topologies (with approximately the same volume)
are compared. Lines are guide to the eyes.}
\end{figure}

\section{Temporal Evolution}
In this section we deal with quantities depending explicitly on time $t$.
First of all, we consider the concentration $\rho_{A}$ of $A$ particles
present at time t. Due to the irreversibility of the reaction taken into
account, $\rho_{A}$ is a monotonic increasing function; more precisely it
is described by a sigmoidal law, typical of autocatalytic phenomena [7].

As shown in Fig.4 the curves $N_{A}(t)$ grow faster, and saturate earlier,
with increasing $\tilde{d}$ ($N$ and $V$ being fixed). This is consistent with the
meaning of the spectral dimension $\tilde{d}$: it describes the long-range
connectivity structure of the substrate and the long-time diffusive
behaviour of a random walker on the substrate. More precisely, for $\tilde{d} <
2$, the number of different sites visited by each walker grows faster as
$\tilde{d}$ increases, and analogously the number of meetings between walkers.

For $\tilde{d} \ge 2$ (e.g., $\tilde{d}=3$ in the figure), $N_{A}(t)$ is
independent of $\tilde{d}$ and is fitted by a pure sigmoidal function. Also
notice that deviations between curves relevant to different topologies are
especially important at early-times, while at long times they all agree with
the pure sigmoidal curve. This result is consistent with the existence of
two temporal regimes concerning diffusion on low-dimensional structures [1].
As a result, the topology of the underlying structure is important only at
early times, while, at long times, the system evolves as expected for
high-dimensional structures.

\begin{figure}[htbp]\label{fig1}
\includegraphics[width=3.0in,height=2.80in]{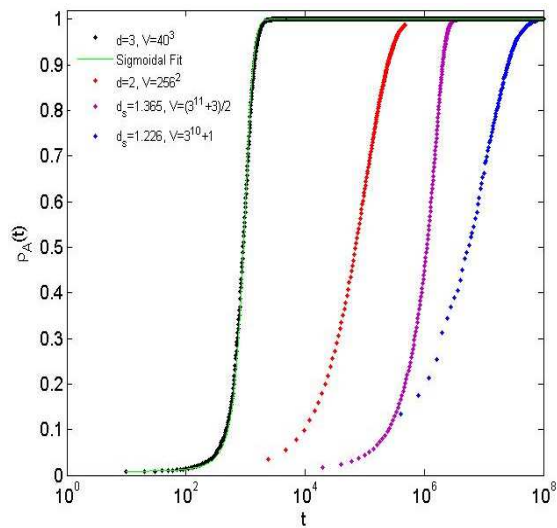}
\caption{Normalized number of $A$ particles $N_{A}(t)/N$ vs time $t$ for a system
made up of $N=128$ particles embedded on different structures, as explained in
the legend. The best fit for the cubic lattice a pure sigmoidal function
(see Eq. (5)), shown by the green line. The latter also provides the best
fit for the long time behaviour of $N_{A}(t)/N$ on low dimensional
substrates.}
\end{figure}

Within the analytic framework developed in the last section, it is possible
to derive some insights into the temporal behaviour displayed by $N_{A}(t)$.
Being $t(n)$ the average time at which the number of $A$ particles reaches value
$n$, recalling Eq. (\ref{eq2}) we can write

From which $N_{A}(t)=t^{-1}(N_{A})$, whose numerical solution provides
an S-shaped curve consistent with data obtained from simulations.

As for transient lattices, the easy form obtained for $P_{\mathrm{trap}}(\rho_{A}, t)\sim pN_{A}$ and the assumption of a uniform distribution for
agents positions, allow to write a Master equation for the number of $A$
particles in the system:
\begin{equation}
N_{A}(t+1) = N_{A}(t) + (N - N_{A}(t)) [1 - (1 - p)^{NA(t)}].
\end{equation}

To first order in $p$: $N_{A}(t+1) =\mathcal{L}_{p} (N_{A}(t))$, being
$\mathcal{L}_{p}$ a logistic-like map, with a repelling fixed point in $0$
$(f'(0)=1+Np)$, and an attracting fixed point in $N$ $(f'(N)=1-Np)$. Since $Np \sim \rho \ll 1$, the increment of $N_{A}(t)$ at each time step is very
small (of order $p$), and we can take the evolution to be continuous. Thus we
obtain
\begin{equation}\label{eq5}
\rho _{A}(t)=e^{Npt} (e^{Npt} + N - 1)^{- 1}
\end{equation}
which is in good agreement with numerical results (Fig.4).

\begin{figure}[htbp]\label{fig1}
\includegraphics[width=3.0in,height=2.80in]{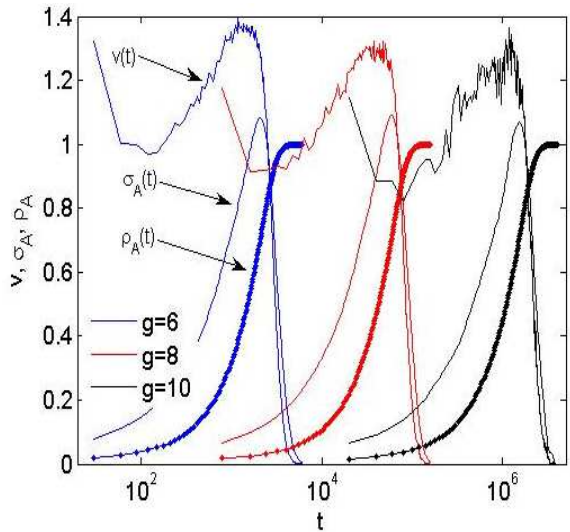}
\caption{Reaction velocity $v$, Fluctuations $\sigma_{A}$ and concentration
$\rho_{A}$ versus time for a system of $N=128$ particles diffusing on a
Sierpinski gasket; three different generations (depicted in different
colours) are shown. Notice $t_{v}<t_{\sigma}$.}
\end{figure}

From $\rho_{A}(t)$ one can derive the rate of reaction $v(t)=\partial_{t}N_{A}(t)$ which represents the reaction velocity. As you can see
from Fig.5, in agreement with the theoretical predictions, $v(t)$ is an
asymmetrical curve exhibiting a maximum at a time denoted as $t_{v,}$
obviously corresponding to a flex in $N_{A}(t)$. Interestingly, $t_{v}$
scales with the volume of the structure according to $t_{v}\sim V^{2 /
\tilde{d}}$ which is the same dependence shown by $\tau$. Moreover, at $t_{v}$ the
population of the two species are about the same
$(N_{A}(t_{v})=N_{B}(t_{v})=N/2)$.

Hence, the efficiency of the autocatalytic reaction is not constant in time
but, provided the number $N$ of particles is conserved, it is maximum when the
number of B particles is about $N/2$. From Eq. we can derive a similar result
for the variance $\sigma_{A}(t)$ of the number of A particles present on
the substrate. Interestingly, fluctuations $\sigma_{A}(t)$ peak at a time
$t_{\sigma}$ which, again, depends on the system size with the same law as
$\tau$; notice that $t_{\sigma} > t_{v}$.

\section{Conclusion}
We introduced an analytic approach to deal with autocatalytic
diffusion-reaction processes, also able to take into account the role played
by particles discreteness and substrate topology. Within such framework, we
derived in the low-density regime, for both fractal and Euclidean
substrates, the exact dependence on system parameters displayed by the
average Final Time, also highlighting how topology affects it. In
particular, the case $d=2$ is marginal. Exact results are also found for
Euclidean lattices in the limit of high density.

Theoretical results concerning the average Final Time find important
applications in analytical fields, where measures of $\tau$ are exploited
for detecting trace reactants. Our results suggest that the sensitivity of
such technique is affected not only by the reactant concentration, but also
by the topology of the structure underlying diffusion.

\section{References}

[1] S. Havlin, D. ben Avraham, Diffusion and Reactions in fractals and
disordered systems, Cambridge University Press, Cambridge, 2000

[2] P.G. de Gennes, J. Chem. Phys. \textbf{76} 3316 (1982)

[3] D. Toussaint, F. Wilczek, J. Chem. Phys. \textbf{78} 2642 (1983)

[4] E. Agliari, R. Burioni, D. Cassi, F.M. Neri, Phys. Rev. E \textbf{73}
046138 (2006)

[5] E. Agliari, R. Burioni, D. Cassi, F.M. Neri, Phys. Rev. E \textbf{75}
021119 (2007)

[6] E. Agliari, R. Burioni, D. Cassi, F.M. Neri, Theor. Chem. Acc. (2007)

[7] J. Mai, I.M. Sokolov, A. Blumen, Europhys. Lett. \textbf{4} 7 (1998)

[8] C.P. Warren, G. Mikus, E. Somfai, L.M. Sander, Phys. Rev. E \textbf{63}
056103 (2001)

[9] R.A. Fisher, Ann. Eugenics \textbf{7} 335 (1937), A. Kolmogorov, I.
Petrovsky, P. Piskunov, Bull. Univ. Moscow. Ser. Int. Sec. A \textbf{1} 1
(1937).

[10] M. Endo, S. Abe, Y. Deguchi, T. Yotsuyanagi, Talenta \textbf{47} 349
(1998)

[11] M. Ishihara, M. Endo, S. Igarashi, T. Yotsuyanagi, Chem. Lett.
\textbf{5} 349 (1995)

[12] K. Ichimura, K. Arimitsu, M. Tahara J. Mater. Chem. \textbf{14} 1164
(2004)

[13] A. Rose, Z. Zhu, C.F. Madigan, T.M. Swager, V. Bulovic Nature,
\textbf{434} 876 (2005); N.D. Priest, J. Environ. Monit. \textbf{6} 375
(2002); J.R. McKeachie, W.R. van der Veer, L.C. Short, R.M. Garnica, M.F.
Appel, T. Benter Analyst, \textbf{126} 1221 (2001)

\end{document}